# Excitonic complexes in MOCVD-grown InGaAs/GaAs quantum dots emitting at telecom wavelengths

Paweł Mrowiński[1,3], Anna Musiał[1,†], Krzysztof Gawarecki[2], Łukasz Dusanowski[1,*], Tobias Heuser[3], Nicole Srocka[3], David Quandt[3], André Strittmatter[3,**], Sven Rodt[3], Stephan Reitzenstein[3], and Grzegorz Sęk[1]

[1]*Laboratory for Optical Spectroscopy of Nanostructures, Department of Experimental Physics, Faculty of Fundamental Problems of Technology, Wrocław University of Science and Technology, Wybrzeże Wyspiańskiego 27, 50-370 Wrocław, Poland*

[2]*Department of Theoretical Physics, Faculty of Fundamental Problems of Technology, Wrocław University of Science and Technology, Wybrzeże Wyspiańskiego 27, 50-370 Wrocław, Poland*

[3]*Institute of Solid State Physics, Technical University of Berlin, Hardenbergstraße 36, D-10623 Berlin, Germany*

*Present address: Technische Physik, University of Würzburg, Am Hubland, D-97074 Würzburg, Germany*

**Present address: Institute of Experimental Physics, Otto von Guericke University Magdeburg, D-39106 Magdeburg, Germany*

Hereby, we present a comprehensive experimental and theoretical study of the electronic structure and optical properties of excitonic complexes in strain-engineered InGaAs/GaAs quantum dots (QDs) grown by metal-organic chemical vapour deposition and emitting at the 1.3-μm telecommunication window. Single QD properties have been determined experimentally for a number of nanostructures by means of excitation-power-dependent and polarization-resolved microphotoluminescence and further compared with the results of confined states calculations employing the 8-band k·p theory combined with the configuration interaction method. The origin of excitonic complexes has been exemplarily confirmed based on magnetooptical and correlation spectroscopy study. Understanding the influence of structural parameters and compositions (of QDs themselves as well as in the neighbouring strain reducing layer) allows to distinguish which of them are crucial to control the emission wavelength to achieve the telecommunication spectral range or to affect binding energies of the fundamental excitonic complexes. The obtained results provide deeper knowledge on control and on limitations of the investigated structures in terms of good spectral isolation of individual optical transitions and the spatial confinement that are crucial in view of QD applications in single-photon sources of high purity at telecom wavelengths.

## I. INTRODUCTION

Semiconductor nanostructures have become a powerful and flexible technology platform to study quantum optics phenomena in the solid state.[1] The main advantage of solid-state structures with quantum dots (QDs) in comparison to atomic systems is their compactness and the possibility to tailor their selected properties for a specific application. To optimize QDs for novel quantum technologies such as, e.g., quantum communication networks, it is crucial to perform comprehensive studies of their optical properties and the underlying electronic structure as well as to identify their determining factors by means of excitonic and excited state spectra, in particular on the level of a single quantum emitter. In case of fiber-based quantum communication, it is of great importance to develop QD-based non-classical light sources emitting in the telecommunication spectral windows at 1.3 and 1.55 µm. One of the approaches to target this challenge is to use GaAs-based QDs (emission typically centered at 900-1050 nm at 10 K for InGaAs/GaAs QDs) and redshift their emission to longer wavelengths accordingly. Various methods have been explored to reach the telecom range in this material system by: (i) engineering the strain in InAs/GaAs QDs utilizing an InGaAs strain reducing layer (SRL)[2–6,7–10] (ii) by using SRL containing Sb for deeper confining potential;[11–14] (iii) deposition of the InAs QDs on InGaAs metamorphic buffer layers;[15,16] (iv) using bilayers of differently-sized QDs acting as the first, seeding layer which modifies the strain conditions for the second one;[17–20] (v) increasing the QD height by growth up to the second critical thickness;[21,22] (vi) increasing the QD size using atomic layer molecular-beam epitaxy (MBE);[23] (vi) using activated alloy phase separation[24] or (vii) nitridation of InAs/GaAs QDs leading to formation of dilute nitride InAsN QDs.[25] In that respect, it is especially desirable to obtain emission around the 1310 nm spectral window due to lack of dispersion and local minimum of losses for the standard fiber networks which is achievable in the GaAs-based structures and suitable for applications in local networks for short- and medium-range communication.

If compared to InP-based approaches[26–29], GaAs-based technology combines the advantage of mature fabrication and material processing with compatibility and feasible on-chip integration. Additionally, the fabrication of high-quality distributed Bragg reflectors (DBRs) to enhance the extraction efficiency of the QD emission as well as microcavities is much less demanding in GaAs-based structures in comparison to InP-based counterparts due to availability of compatible materials with high refractive index contrast. Furthermore, in the case of InAs on InP system it is rather optimal for longer wavelengths of the third window at 1550 nm and covers the range of absolute minimum of losses preferential for long-haul data transmission, but on the expense of possible distortion of the optical signal due to dispersion.

In this paper we investigate fundamental physical properties of excitonic complexes confined in InGaAs/GaAs QD structures capped with InGaAs strain reducing layer of lower In content, in view of their practical implementation in non-classical light sources for short range quantum communication protocols at the telecom O-band.[30] The quality of these structures have reached the level enabling systematic experimental study on many QDs, but their practical application still requires optimization as the structural quality and therefore the internal quantum efficiency of the emitters is not yet comparable to their counterparts emitting at shorter wavelengths. Ensembles of such structures with high spatial density have already been investigated in view of laser applications focusing on the role of the SRL in redshifting the QDs' emission towards telecommunication wavelengths, however on the level of the averaged optical response of the entire ensemble.[5,31,32] Here, we focus on the single quantum emitters and on identifying their optical and electronic properties including the exciton fine structure splitting – bright and dark excitonic states – as well as the binding energies of excitonic complexes as a function of emission energy and morphological parameters. This enables us to point out efficient single optical transitions of good thermal stability. We study both experimentally and theoretically the properties of various excitonic complexes by means of microphotoluminescence (µPL), also in magnetic field, and compare them with 8-band k·p modelling[33] combined with the configuration interaction method proven to reflect properly the excitonic states in different types of epitaxial QDs.[34,35] The analysis of the single-particle states and the binding energies of excitonic complexes with respect to such system parameters as QD geometry, QD composition and SRL composition is necessary for proper understanding of the interdependence between the structural, electronic and optical properties of single QDs grown by metal organic chemical vapor deposition (MOCVD), being relatively cheap and efficient technique compared to other fabrication methods. Moreover, it provides means for tailoring the dots characteristics to make them suitable for specific nanophotonic applications in the O-band fiber window. So far, similar QD structures with however different structural properties and ultra-low spatial density were studied with respect to cascaded emission of photons (from both neutral and charged excitonic complexes)[10] and temperature dependence of photoluminescence[36] followed by demonstration of single-photon emission.[37] However, these structures exhibit mainly negative charged complexes, whereas in our case the presence of both positively (X+) and negatively (X-) charged trions are present in the spectra. Their relative intensity differing from dot to dot suggests that random local charge environment is mostly responsible for the charge state of individual QDs. There is also a theoretical work on the fine structure splitting of the neutral exciton in similar family of InAs/GaAs with the SRL,[7] but neither the binding energies of the multicarrier excitonic complexes, nor their interplay with the ground state emission energy or energetic order of the respective excitonic complexes have been studied. Therefore, our aim is to fill this gap. In addition, we also extend the existing knowledge by direct detection of the dark excitonic states in these nanostructures which has not been reported before. This is an analysis which is fully complementary to the recent comprehensive study on structural and optical properties of GaAs-based QDs on metamorphic buffer layer for emission at longer wavelengths of the telecom C-band[38], for which emission of single photons and entangled photon pairs has been demonstrated.[39,40]

The paper is organized as follows: Section II provides details on the investigated structures together with a description of the utilized experimental setups. In Section III experimental findings are discussed, whereas Section IV presents the quantum dot model used in theoretical calculations together with the obtained results and their comparison to experiments. Section V concludes the paper.

## II. INVESTIGATED STRUCTURES AND EXPERIMENTAL SETUP

The investigated epitaxial heterostructure containing self-assembled $In_xGa_{1-x}As$ on GaAs QDs was grown by MOCVD in the Stranski-Krastanov growth mode. QDs on a wetting layer are formed of $In_{0.75}Ga_{0.25}As$ (0.7 nm of the nominal deposited material). The QDs of ~ $10^9/cm^2$ areal density are covered by a low-indium content $In_{0.2}Ga_{0.8}As$ SRL, purpose of which is to redshift the ground state transition energy to the telecom spectral range.[31,41] For enhanced extraction efficiency of the emitted radiation, QDs are grown on a distributed Bragg reflector (DBR) composed of 23 pairs of GaAs/AlGaAs layers (101.6/115.4 nm thicknesses measured by scanning electron microscopy - SEM) on top of the undoped GaAs buffer (300 nm). The thickness of the GaAs layer surrounding the QD layer is designed to form a 2-λ cavity between the DBR and the sample surface [Fig 1(a)]. Additionally, cylindrical mesa structures of various diameters in the range of 300 nm to 2100 nm and 670 nm height are fabricated in a regular pattern using electron-beam lithography performed at room temperature followed by reactive-ion etching (ICP-RIE) [inset in Fig. 1(b)]. Such mesas provide increased spatial resolution by removing neighbouring QDs and hence also their possible contribution to the background emission (see an exemplary μPL spectrum in Fig. 1(b)]. Fabricating such mesas also improves the directionality of the QD emission - it has been shown theoretically that optimized mesa design can lead to extraction efficiencies comparable to those achievable with microlenses, but it is more robust against fabrication imperfections[42] which has also been confirmed experimentally within deterministic technology platforms.[43,44]

All the presented µPL results are obtained for the described sample mounted in a liquid helium continuous-flow cryostat (at 5 K) and under non-resonant continuous wave (cw) excitation at 660 nm. The excitation is delivered to the sample and the signal is collected via a long working distance microscope objective with 0.4 numerical aperture. The signal is further spectrally resolved by a 1-m focal length monochromator with a 600 grooves/mm grating and detected using a multichannel liquid nitrogen-cooled InGaAs linear detector. The setup provides a spectral resolution of at least 25 µeV and spatial resolution of about 1 µm. For polarization-resolved measurements, we used a rotating half-wave plate and a linear polarizer adjusted for the maximal transmission of the experimental setup. The magnetooptical study was performed using a microscopy cryostat with superconducting coils generating magnetic fields up to 5 T which allows for measurements in both, Voigt and Faraday configurations. For photon auto- and cross-correlation measurements, we utilized a free-space Hanbury Brown and Twiss (HBT) setup with a non-polarizing 50:50 beam splitter and two 0.32 m focal length monochromators used as spectral filters and a pair of superconducting NbN nanowire single-photon counting detectors with ~20% quantum efficiency and 10 dark counts/s at 1.3 µm. The photon correlation statistics was acquired by single-photon correlation electronics providing an overall temporal resolution of the HBT setup of 80 ps.

## III. EXPERIMENTAL RESULTS AND DISCUSSION

The optical properties of all investigated QDs have been examined by performing high-resolution µPL experiments as a function of excitation power and linear polarization-resolved. Based on this the basic excitonic complexes have been identified for more than 10 QDs and allowed to determine interdependences between the optical properties of investigated structures and further, thanks to comparison with calculations, relate them to the structural parameters. For an exemplary typical QD also magnetooptical and cross-correlation measurements have been performed (see next paragraphs), which confirmed the initial identification of excitonic complexes based on µPL study. Characteristic spectral pattern (energetic order) of the basic excitonic complexes can be determined from the optical study and is reproduced by the results of the modelling which additionally brings an argument to distinguish between positively and negatively charged excitons. The representative emission spectrum indicating a set of single QD lines from a circular mesa structure of 1300 nm in diameter is shown in Fig 2(a). It presents two spectra measured for polarization along the [110] and [1-10] in-plane crystallographic directions. Based on the excitation power dependence [Fig 2(b)], the full scan of polarization angle [Fig 3(a)] and photon cross-correlation measurements [Fig 5(a)] one can unambiguously identify the neutral exciton (X) and the biexciton (XX) by: i) the opposite fine structure splitting (FSS) for XX and X in terms of the energetic order of the polarized components; ii) the expected approximately linear (or sublinear) and quadratic excitation power dependence of the emission intensity in the low excitation limit for X and XX, respectively; iii) the anti-bunching followed by bunching being a signature of cascaded emission in the cross-correlation of photon-emission events from X and XX [Fig. 5(a)]. On the other hand, i) the lack of the spectral splitting in the polarization-resolved study [Fig. 3(a)], ii) the sublinear emission intensity dependence on excitation power [Fig. 2(b)] and iii) a quadruplet splitting observed in the in-plane magnetic field [Fig. 3(b) and Fig. 4(a)], all observed only for two QD emission lines indicate their charged exciton character (labelled as $X^+$, $X^-$ in Fig. 2). Additionally, the cross-correlation measurements between the charged and neutral exciton (example for $X^+$ is shown in [Fig. 5(c)]) prove unambiguously that all these identified emission lines originate from the same single QD. The argument for assigning the signs of the charged excitons is brought by the calculations presented below [Fig. 11] which indicate three times lower binding energy of the positive trion as compared to the negatively charged one in the experimentally relevant parameter range. Additionally, the emission lines identified as X+, X- and X are the ones visible in the spectrum in the low excitation regime (not shown here) which confirms they are rather basic excitonic complexes in the ground state due to low probability of forming complex carrier configurations including higher energy states in this excitation conditions. The very characteristic behavior of XX described above and linked to the properties of X allows to distinguish it from other emission lines appearing in the spectrum with increasing excitation power (Fig. 2(a)), which are not of interest here and cannot be reliably identified. They could be related either to the excited states or higher-order excitonic complexes including carriers in the p-shell form the same QD as reported previously for more common InAs/GaAs system[47,48] or emission from a different QD located in the same mesa. The identification of all the emission lines in the presented spectrum is beyond the scope of this work as we focus on the basic excitonic complexes which can be reliably analyzed within our approach, and it does not influence the generality of the obtained results neither the related conclusions on the fundamental optical properties of the investigated structures.

Additionally, for the X line an auto-correlation of photon emission events under non-resonant cw excitation was measured [Fig. 5(b)]. The second-order correlation function at zero time delay gives the as-measured value of $g^{(2)}(0) = 0.17$, and $g^{(2)}(0) = 0.05$ after deconvolution with temporal response of the experimental setup (80 ps), which proves the single-photon character of the emission. The non-ideal value is attributed to uncorrelated background emission either due to multiple QDs in a mesa for this spatial QD density or due to emission from defects in the structure (most probably in the substrate). In this experiment, the fit is performed using the expression $1 - (1 - g^2(0))e^{-\frac{|t|}{t_c}}$ with recovery time constant $t_c$ of about 0.7 ns, which is close to the typical excitonic lifetime for investigated QDs on the order of 1 ns. In the case of a trion-exciton cross-correlation measurement shown in Fig. 5(c), the dip is broader and asymmetric with characteristic recovery time constants equal to: 0.9 ns for $\tau < 0$ (similar to the time scale from X-X autocorrelation measurements) and 2 ns for $\tau > 0$. The excitation power is slightly lower than in the case of X-X autocorrelation and therefore the dip is expected to be broader. However, the substantially longer timescale for positive time delays should rather be related to the time needed for the X+ to be formed after recombination of X than to the radiative lifetime. Recombination of X leaves the QD empty and 2 ns is apparently required to capture two holes and one electron in the QD and to form the X+ (the excessive carrier can origin from a trap or defect state which makes the process slower).

The identification of excitonic complexes of the single QD under study allows us to determine their binding energies to be: $\Delta E_{X^+}$ = -1.7 meV, $\Delta E_{X^-}$ = -3.5 meV and $\Delta E_{XX}$ = -3.7 meV, for, $X^+$, $X^-$ and XX, respectively, being in the range of values similar to MBE-grown QDs[35,49] and low-density MOVPE-grown QDs[10] in this material system, but emitting below 1 μm wavelength. The FSS of the neutral exciton (60 μeV) is also rather typical for In(Ga)As/GaAs (001) QDs emitting below 1 μm wavelength,[50] which were not especially optimized for low FSS. In order to obtain the splittings of the bright-dark and dark-dark exciton spin configurations, diamagnetic coefficients as well as carrier and exciton g-factors we collected also the magneto-optical data. In Figure. 3(b) and 3(d) we demonstrate PL spectra in Farady and Voigt configuration from 0 to 5 Tesla, respectively. Figure 3(c) and 3(e) shows the PL peak energy corresponding to various excitonic complexes influenced by an external magnetic field up to 5 T together with the emission spectra at 0 and 5 Tesla. The grey lines represent fit to match the experimentally obtained dependencies, i. e., emission energy of the respective emission lines vs. magnetic field after the model adapted from ref.[51], where the complete description of the spin splitting in external magnetic field for in-plane asymmetric quantum dot can be found. The model includes the electron-hole exchange interaction and the interaction of electron and hole spins with an external magnetic field (Zeeman effect). Using the exciton states basis ($|+1\rangle, |-1\rangle, |+2\rangle, |-2\rangle$), the Hamiltonian of the system can be represented in a form of:

$$H = H_{exchange} + H_{Zeeman} = \frac{1}{2}\begin{pmatrix} \delta_0 & \delta_1 & 0 & 0 \\ \delta_1 & \delta_0 & 0 & 0 \\ 0 & 0 & -\delta_0 & \delta_2 \\ 0 & 0 & \delta_2 & -\delta_0 \end{pmatrix} + \frac{\mu_B B}{2}\begin{pmatrix} g_X & 0 & g_{e,\perp} & g_{h,\perp} \\ 0 & -g_X & g_{h,\perp} & g_{e,\perp} \\ g_{e,\perp} & g_{h,\perp} & -g_X + 2g_h & 0 \\ g_{h,\perp} & g_{e,\perp} & 0 & g_X - 2g_h \end{pmatrix},$$

where $\delta_{0,1,2}$ are bright-dark, bright-bright and dark-dark splitting of exciton state, $g_X = g_e + g_h$ is exciton g-factor, $g_e$ ($g_h$) is perpendicular electron (hole) g-factor and $g_{e,\perp}$ ($g_{h,\perp}$) is in-plane electron (hole) g-factor. In addition, the effect of exciton diamagnetic shift is included as proportional to $aB^2$ in the low-field limit (energy shift is small compared to exciton binding energy), where a describes the diamagnetic coefficient.[52] In the case of biexciton state, which is a singlet state, the diamagnetic coefficient is defined as $a_{XX} \cong 2a_X$ and the Zeeman splitting observed in the experiment is due to the splitting in the final state of recombination process, which is the exciton state. Therefore, the magnetic field dependence for XX transition is the same as for X shifted by the XX binding energy. In order to account for the magnetic field dependence for charged excitonic complexes one can expect the same Zeeman splitting as for X, shifted by the X+ (X-) binding energy, as the origin of the splitting is proportional also to the sum of $g_e$ ($g_h$) in the initial X+ (X-) state and $g_h$ ($g_e$) in the final hole (electron) state.[53] The parameters of the fitting curve are determined from the uPL measurements without external magnetic field, i. e. the zero-field emission energy of the excitonic state ($E_{0X}$,), binding energies of the basic excitonic complexes ($\Delta E_{X+}, \Delta E_{X-}, \Delta E_{XX}$) and the zero-field X FSS ($\delta_0$). The remaining parameters are fitting parameters ($\delta_{1,2}, g_X, g_{e,\perp}, g_{h,\perp}, a_X$) in order to obtain minimum deviation from experimental data.

Figure 4(a) schematically explains the origin of the observed Zeeman splittings. Experiments in Voigt configuration allow to observe the dark excitonic states due to their mixing with bright excitonic states. By extrapolating the energies of the respective excitonic components down to zero magnetic field, one can determine both the bright-dark and dark-dark X splittings. This procedure enables us to identify the zero-field energy of the dark states, and hence get the bright-dark splitting of 430 μeV and the dark-dark splitting of approx. 20 μeV. Additionally, the quadruplet splitting of the charged excitons allows determining the absolute values of the in-plane g-factors of electron and hole yielding $|g_{e,\perp}| = 0.99$ and $|g_{h,\perp}| = 0.28$. A non-zero in-plane hole g-factor suggests slight contribution of the light-hole states to the valence band ground state, which can result from an intrinsic anisotropy of the confining potential due to, e.g., the QD shape anisotropy or inhomogeneous strain around the QD. The presence of the valence-band mixing is also seen in the non-zero polarization anisotropy of the exciton emission of about 5% in agreement with rather low asymmetry of the in-plane QD shape deduced from structural data (not shown here). The diamagnetic shift in Voigt configuration is characterized by the coefficient $a_{X,\perp} = 2\ \mu eV/T^2$. Experiments in Faraday configuration yield the Zeeman splitting characterized by the exciton g-factor $g_X = 1.75$ and diamagnetic coefficient of $a_X = 13\ \mu eV/T^2$.

The results discussed above from magnetooptics and correlation spectroscopy confirmed that the initial identification of basic excitonic complexes based on excitation-power-dependent and polarization-resolved μPL is correct and sufficient. Therefore, in the next step, we performed the same analysis as the exemplary one presented in Figs. 2 and 3(a) for several QDs, namely the identification of different excitonic complexes was based solely on the μPL results (a similar spectral pattern, FSS and excitation power dependence of the emission intensity). This enabled to gather larger statistics and to determine the interdependences of their optical properties. The results concerning X emission energy dependence of the exciton FSS and the binding energies of the excitonic complexes are presented in Fig. 6. The histogram of exciton FSS [inset in Fig. 6(a)] shows a statistical distribution from 30 μeV to 85 μeV with a pronounced maximum at 65 μeV. At the same time, there is no clear X energy dependence of the exciton FSS probably due to several factors contributing to the exciton energy[54] - see the discussion below. Furthermore, we observe no clear dominance of emission intensity of any of the charged excitons. The presence of both X+ and X- in the spectra at low excitation power regime suggests that random local charge environment differing from dot to dot is mostly responsible for the relative intensity of the opposite charge trion states. The charge state of the QD depends also on the details of the growth procedure determining the structural material quality and the type and level of unintentional doping, not only the growth method itself as, in contrast to our findings, clear dominance of negatively charged complexes reported for similar QDs in Ref. 10. In the case of the binding energies of basic excitonic complexes [Fig. 6(b)], we observe a large scattering for the XX and X- (with however some general trends indicated by the guide to the eye) and a rather weak dependence as a function of the X energy, whereas a clear decrease of the binding energy of the X+ complex with increasing exciton energy is observed. The scattering of the data for the XX and X- binding energy is probably due to less localized electron wave function as compared to the hole counterpart making these binding energies more sensitive to changes in In composition and its distribution influencing simultaneously the band-gap potential and the effective masses, and thus direct Coulomb interactions and correlation effects.[35] At the same time, a rather smooth X+ binding energy dependence can be related to a more localized hole wave function [see also Fig. 9(b)] and according to the results of k·p modelling (see details below) its character (mathematical trend) is most probably driven by the In content. Perhaps, the same fluctuations of the confinement potential being related to the QD composition could be also responsible for the FSS distribution.[55,56]

## IV. THEORETICAL MODEL AND RESULTS

In this section, a model of the QD as well as details of the methods utilized to calculate the electronic/excitonic structure are presented. We assume realistic shape of the dot[57] by describing its upper surface as:

$$Z(x,y) = h\exp\left[-\left(\frac{x^2+y^2}{r_b^2}\right)^2\right], \quad (1)$$

where $h$ denotes the height and $r_b$ is an in-plane extension parameter related to the base size. The dot is placed on a $h_{WL}$ thick $In_{0.75}Ga_{0.25}As$ wetting layer (WL). The material distribution in the system is shown in Fig. 7(a) for our model QD design. Fig. 7(b) shows a transmission electron microscopy (TEM) cross section of a representative QD. There is presented the spatial mapping of the real part of the interference between the {000} and {200} (in-plane direction) beams. Within the approximate specimen thickness of ~50 nm the colored scale represents the projected In concentration along the electron beam thus given a qualitative impression of the QD shape. However, from this TEM data it is not possible to deduce the specimen thickness and thus the absolute value of the In content within a certain cross-sectional plane. Based on that image we assume a gradient material distribution inside the QD. To determine the In content the sample was investigated under a 21.7° out of the [010] plane rotation angle (not shown here). In that case, the sample thickness can be estimated from reflex positions of the WL, the QDs are modeled as hemispheres and the In content can be calculated. The selected QD has the In content of (77±13) %. Therefore, in the QD model the local In content $C(x,y,z)$ changes from $C_{max}$ to $C_{min}$ within the assumed QD shape, i.e. $0 \leq z \leq Z(x,y)$, according to the dependence

$$C(x,y,z) = C_{min} + (C_{max} - C_{min})\exp\left(-\frac{x^2+y^2}{r_0^2}\right)\exp\left(-\frac{(z-z_c)^2}{z_0^2}\right), \quad (2)$$

here $r_0$, $z_0$ define spatial extensions and $z_c$ is related to the position of the gradient. The parameter set for a representative QD (optimized according to the experimental results on the QD ensemble – average values) is given by: $h$ = 6 nm, $r_b$ = 15 nm, $h_{WL}$ = 0.6 nm, $C_{max}$ = 0.95, $C_{min}$ = 0.5, $r_0$ = 12 nm, $z_0$ = 6 nm, $z_c$ = $h/2$. This gives an average In content inside the dot of $C_{avg}$ = 0.727 similar to the one obtained from the structural data. The strain reducing layer far aside the QD has a constant thickness equal to the nominal value of $h_{SRL}$ = 4.2 nm and In content $C_{SRL}$ = 0.2. On the other hand, in the QD vicinity, we assume its upper surface is given by Eq. 1 with h → h'$_{SRL}$ = 7.8 nm, $r_b$ → r'$_{SRL}$ = 30 nm. The lattice mismatch between InGaAs and GaAs results in a strain field, which affects the band structure. We calculate the strain distribution in the system within a continuous elasticity approach.[58] The piezoelectric field is accounted for by using a strain-induced polarization up to the second order.[59] The electron and hole single-particle states are calculated within the 8-band k·p method.[60] Further details of the calculations as well as the material parameters are presented in Ref. [[61]].

The total Hamiltonian of the system in the picture of second quantization is given by

$$H = \sum_n \epsilon_n^{(e)} a_n^\dagger a_n + \sum_m \epsilon_m^{(h)} h_m^\dagger h_m - \sum_{nn'mm'} V_{nmm'n'}^{(e-h)} a_n^\dagger h_m^\dagger h_{m'} a_{n'} + \frac{1}{2}\sum_{nn'mm'} V_{nn'mm'}^{(e-e)} a_n^\dagger a_m^\dagger a_{m'} a_{n'} + \frac{1}{2}\sum_{nn'mm'} V_{nm\,n'}^{(h-h)} h_n^\dagger h_m^\dagger h_{m'} h_{n'},$$

where $\epsilon_n^{(e)}$ ($\epsilon_m^{(h)}$) are the energies of the electron (hole) states found from 8-band k·p calculations, $a_n^\dagger$, $a_n$ ($h_m^\dagger$, $h_m$) are operators of creation and annihilation of electrons (holes) in state $n$ ($m$). The Coulomb integrals are defined by

$$V_{nmm'n'}^{(e-h)} = \frac{e^2}{4\pi\epsilon_0\epsilon_r}\int d\mathbf{r}\int d\mathbf{r'}\,\Psi_n^{\dagger(e)}(\mathbf{r})\Psi_m^{\dagger(h)}(\mathbf{r'})\frac{1}{|\mathbf{r}-\mathbf{r'}|}\Psi_{m'}^{(h)}(\mathbf{r'})\Psi_{n'}^{(e)}(\mathbf{r}),$$

where $e$ denotes the electron charge and $\epsilon_0$, $\epsilon_r$ are vacuum perimitivity and relative dielectric constant for GaAs. $\Psi_n^{(e)}(\mathbf{r})$ ($\Psi_m^{(h)}(\mathbf{r'})$) are 8-component spinors related to the electron (hole) eigenstates. An analogous definition is used in the case of $V_{nmm'n'}^{(e-e)}$ and $V_{nm\,n'}^{(h-h)}$.

To find exciton, trion and biexciton states we use the configuration interaction approach. We take into account a basis of 40 electron and hole states (including spin degeneracy) which gives 1 600 exciton, 31 200 trion and 608 400 biexciton configurations. The resulting matrices for trions and biexciton are sparse and they are diagonalized using the SLEPC library.[62] The importance of the correlations in the investigated system can be illustrated by showing the binding energies of excitonic complexes as a function of the number of states included in the calculations [Fig. 8(a)]. Convergence tests reveal that the correct description of long-wavelength InGaAs/GaAs QDs indeed requires modelling beyond the Hartree-Fock approximation to properly describe the excitonic states of the investigated system.

The emission spectrum of our model QD was calculated and compared to the experimental results presented above, also in order to verify the assumed system parameters. In the first step, a single-particle energy level spectrum [Fig. 9(a)] and wave functions [Fig. 9(b)] were calculated. The in-plane carrier density probability distribution has a high symmetry as the QD shape and In composition with cylindrical symmetry (assumed following the low degree of linear polarization of exciton emission observed in the optical experiments). Due to a strongly inhomogeneous In distribution in the QD volume [Fig. 7], the wave function extension is smaller than the physical size of the QD and so the confinement is actually effectively stronger than one might expect from the dimensions of the QD itself. Figure 10 shows that the fundamental properties of the investigated system are well reproduced by the calculations. In Figure 10(a) high-excitation spectra of the QD ensemble, exhibiting emission also from higher QD energy states (upper panel), are compared to the calculated excitonic spectrum (lower panel): the splitting between the ground state and the first higher energy state (p-shell) equals to 75 meV in both experiment and theory. Figure 10(b) presents the binding energies of basic excitonic complexes: calculated (black solid line for QD *$r_b$ = 15 nm, h = 6 nm*, In content in SRL of 20 %, average In content in QD of 73 %, X energy - 0.95708 meV) and determined experimentally (red arrows, X energy 0.95719 eV – Fig. 6(b)). For the sake of this comparison, an experimental example with matching exciton emission energy was chosen, and furthermore the exciton energy was subtracted from both experimental and calculated data. The calculated binding energies of the excitonic complexes are: -0.98 meV, -1.97 meV and -2.29 meV for $X^+$, XX and $X^-$, respectively. The ordering of the excitonic complexes in the spectrum and their relative binding energies are in good agreement with the experimentally obtained values with a systematic shift (underestimating the binding energies) due to the fact that exchange interaction correction is not included in these calculations and the parameters were chosen to reproduce averaged properties of the QD ensemble.

When regarding the experimental data, we do not have access to the exact structural information for an individual QD studied optically, but we do have the spectral distribution of the main properties measured for various dots. It is then possible to determine the interdependence of QDs' optical properties as a function of X emission energy, but there might be multiple interdependent factors causing the change of X energy. Because of that a detailed study was performed exploring the influence of the QD's size and the average In composition in the QD and SRL on the binding energies of the excitonic complexes (Fig. 11). Tailoring the QD size in even broad range gives an almost insensitive X+ binding energy of the size (exciton energy) and a strong increase of the X- and XX binding energy absolute value by about 400 μeV [Fig. 11(a)] - both do not reflect the data of Fig. 6 at all.

On the other hand, the In content in the QD has a strong impact on the binding energies of the trions and it leads to their crossing for low values of average In content and hence to reordering of the exitonic complexes appearing in the emission spectrum [Fig. 11(b)]. The binding energy of the positive trion (X$^+$) decreases (i.e. its absolute value) from -2.4 meV for low In content to -0.91 meV for high In content evidencing that it is a crucial parameter for obtaining isolated emission lines from single QDs also at longer wavelengths. At the same time, when looking at the calculated impact of the In content in the QD on the X+ binding energy, one immediately sees that to reproduce the experimentally observed variation in X+ binding energy the In content within the QDs should change in the range of ± 3 % [see the patterned region in Fig. 11(b)] which is realistic taking into account both the resulting width of the ground state ensemble emission exceeding 50 meV[41] and the QD emission energy range. Additionally, a crossing of negative trion (X$^-$) and biexciton (XX) binding energies can be found in these results showing that the In content fluctuations can lead to reordering of these lines, which is consistent with the experimental results discussed in Sec. III. At the same time, the average QD composition strongly influences the ground state energy, and a high In content is indispensable to maintain the emission in the telecom range and to reproduce the experimental results.

One might try to get an insight into the sensitivity of the binding energies of different excitonic complexes to average In content in the QD while examining the Coulomb interaction matrix elements (their dependence on the average In content in the QD is presented in the Figure 8(b)). The $V_{hh}$ element important in the case of positive trion exhibits much stronger dependence on the average In composition in the QD (larger slope) in comparison to $V_{ee}$ and $V_{eh}$ elements and thus as a result a stronger change of the binding energy of the X$^+$ with composition. On the opposite side, the $V_{ee}$ element driving the binding energy of the X$^-$ shows the smallest changes with the In content in the QD. The opposite trends are related to the relative magnitude of the $V_{hh}$ and $V_{ee}$ elements (for X$^+$ and X$^-$, respectively) with respect to the $V_{eh}$ element and their In content-driven changes lead to reversing the order. Those differences are mainly driven by the depth of the confining potentials for the carriers with opposite charge sign. Holes experience shallower confining potential and are therefore more sensitive to changes of the In content in the QD.

The same is true for the In composition of the SRL, i.e. it mainly influences the ground state energy, but has rather weak impact on the binding energies of the excitonic complexes in the experimentally relevant range [see the patterned region in Fig. 11(c) covering the In content range of 0.15 – 0.25]. If one would like to get the experimental trends just by tuning solely the In content in the SRL, it would require a change in unrealistically broad range (between 0 to 0.25 of indium fluctuations) to achieve 0.35 meV variation for X+ with respect to the experiment – such variation is rather impossible for a 2D-like layer of InGaAs (a few percent In content fluctuation is expected at most).When a QD structure with and without SRL [i.e. its In content equal to 0.0 in Fig. 11(c)] is compared, one might conclude that the presence of the strain reducing layer influences mainly the positively charged exciton complex and the ground state energy, while the biexciton and negatively charged exciton binding energies remain almost constant. The thickness of the SRL has even less impact – when its thickness is large enough to provide X emission energy in the target spectral range (in our case above 3 nm) the binding energies of excitonic complexes are independent of this system parameters (results of the calculations not shown here). Moreover, further increase in the SRL thickness do not shift the X emission energy substantially – the dependence saturates. Thus, our theoretical study on the influence of the QD geometry as well as the average In content in both QDs and SRL shows that experimental results seen in Fig. 6(b) can be well reproduced mainly by tailoring the In content in SRL and QD rather than by In content in SRL or by QD geometry, which have most probably lower impact. Similarly, in the experiment, the QDs with both $X^-$ binding energy higher and lower than the XX binding energy were observed, as shown in Fig. 6(b), evidencing that we are in the regime of their intercrossing. Comparison with theoretical calculations [Fig. 11(b)] proves that these effects are governed and rather sensitive to the average In composition in the QD. It would not be possible to reproduce the experimental trends in binding energies neither by the QD size, nor In composition of the SRL. Furthermore, we observe that smaller QDs with higher In content would be beneficial for the single-photon emitter applications (well-isolated optical transitions at the telecom wavelengths), and would allow one to reach binding energies of the positive trion of up to -1.0 meV while keeping the emission wavelength within the 2$^{nd}$ telecommunication band.

## V. CONCLUSIONS

In this work, we examine experimentally and theoretically the electronic structure and optical properties of single strain-engineered InGaAs/GaAs QDs emitting in the telecom O-band at 1.3 μm. The investigated structures exhibit neutral exciton fine structure splitting of 65 μeV on average and biexciton binding energies similar to common InGaAs/GaAs QDs emitting below 1 μm on the level of -3.5 meV. All excitonic complexes have a binding character with the binding energy of the positive trion in the range of -1 meV and that of the negative one around -3.6 meV. Both the higher energy excitonic spectrum (s-p splitting of 75 meV) as well as the binding energies of excitonic complexes are well reproduced by the 8-band k·p calculations combined with the configuration interaction method. Good quantitative agreement between the results of the experiment and the calculations has been obtained by using a realistic, strongly inhomogeneous In distribution within the QD following the structural data. Additionally, the number of electron and hole basis states constituting the input for configuration interaction calculations has been varied proving the importance of correlation effects in the investigated system - the best results are obtained when 40 electron and 40 hole states are included. Whereas the In composition in both, the QD and the strain reducing layer, has strong impact on the exciton ground state transition energy and leads to the energy redshift towards the telecom range, neither the exact QD size nor the composition of the strain reducing layer (in the realistic range of changes) are able to reproduce properly the experimentally obtained trends in dependence of binding energies of the excitonic complexes vs the exciton emission energy. These, however, can be well reflected by slight differences in the In content of the QD itself, and with its increase the ordering of the excitonic complexes changes substantially in the emission spectra: the positive trion becomes less bound than the negative one, what makes the negative trion bound even stronger than the biexciton. As a result, the In composition in QD mainly determines the properties of excitonic complexes in these MOCVD-grown InGaAs/GaAs QDs, but the balance between the parameters of the QD and strain reducing layer is important to maintain the ground state emission energy in the telecom O-band.

## ACKNOWLEDGEMENTS

We thank Nora Schmitt and Prof. Michael Lehmann (Technical University of Berlin) for the TEM measurements. This work was funded by the FI-SEQUR project jointly financed by the European Regional Development Fund (EFRE) of the European Union in the framework of the programme to promote research, innovation and technologies (Pro FIT) in Germany, and the National Centre for Research and Development in Poland within the 2nd Poland-Berlin Photonics Programme, grant No. 2/POLBER-2/2016 (project value 2 089 498 PLN). Support from the German Science Foundation via CRC 787 and the Polish National Agency for Academic Exchange is also acknowledged. P. Mrowiński gratefully acknowledges the financial support from the Polish Ministry of Science and Higher Education within "Mobilność Plus" programme.

**REFERENCES**

[1] F. Jahnke, *Quantum Optics with Semiconductor Nanostructures* (2012).

[2] K. Nishi, H. Saito, S. Sugou, and J.S. Lee, Appl. Phys. Lett. **74**, 1111 (1999).

[3] V.M. Ustinov, N. a. Maleev, a. E. Zhukov, a. R. Kovsh, a. Y. Egorov, a. V. Lunev, B. V. Volovik, I.L. Krestnikov, Y.G. Musikhin, N. a Bert, P.S. Kop'ev, Z.I. Alferov, N.N. Ledentsov, and D. Bimberg, Appl. Phys. Lett. **74**, 2815 (1999).

[4] J. Bloch, J. Shah, W.S. Hobson, J. Lopata, and S.N.G. Chu, Appl. Phys. Lett. **75**, 2199 (1999).

[5] L. Seravalli, M. Minelli, P. Frigeri, P. Allegri, V. Avanzini, and S. Franchi, Appl. Phys. Lett. **82**, 2341 (2003).

[6] E.C. Le Ru, P. Howe, T.S. Jones, and R. Murray, Phys. Status Solidi C Conf. **1224**, 1221 (2003).

[7] E. Goldmann, S. Barthel, M. Florian, K. Schuh, and F. Jahnke, Appl. Phys. Lett. **103**, 242102 (2013).

[8] M.B. Ward, M.C. Dean, R.M. Stevenson, A.J. Bennett, D.J.P. Ellis, K. Cooper, I. Farrer, C. a Nicoll, D.A. Ritchie, and A.J. Shields, Nat. Commun. **5**, 3316 (2014).

[9] S. Maier, K. Berschneider, T. Steinl, A. Forchel, S. Höfling, C. Schneider, and M. Kamp, Semicond. Sci. Technol. **29**, 052001 (2014).

[10] J. Kettler, M. Paul, F. Olbrich, K. Zeuner, M. Jetter, P. Michler, M. Florian, C. Carmesin, and F. Jahnke, Phys. Rev. B **94**, 045303 (2016).

[11] A. Salhi, S. Alshaibani, B. Ilahi, M. Alhamdan, A. Alyamani, H. Albrithen, and M. El-Desouki, J. Alloys Compd. **714**, 331 (2017).

[12] W.S. Liu, D.M.T. Kuo, J.I. Chyi, W.Y. Chen, H.S. Chang, and T.M. Hsu, Appl. Phys. Lett. **89**, 243103 (2006).

[13] K. Akahane, N. Yamamoto, S. Gozu, A. Ueta, and N. Ohtani, Phys. E Low-Dimensional Syst. Nanostructures **32**, 81 (2006).

[14] A. Hospodková, J. Pangrác, J. Vyskočil, M. Zíková, J. Oswald, P. Komninou, and E. Hulicius, J. Cryst. Growth **414**, 156 (2015).

[15] N.N. Ledentsov, A.R. Kovsh, A.E. Zhukov, N.A. Maleev, S.S. Mikhrin, A.P. Vasil'ev, E.S. Semenova, M. V Maximov, Y.M. Shernyakov, N. V Kryzhanovskaya, V.M. Ustinov, and D. Bimberg, Electron. Lett. **39**, 1126 (2003).

[16] L. Seravalli, G. Trevisi, P. Frigeri, D. Rivas, G. Muñoz-Matutano, I. Suárez, B. Alén, J. Canet, and J.P. Martínez-Pastor, Appl. Phys. Lett. **98**, 173112 (2011).

[17] K. Shimomura and I. Kamiya, Appl. Phys. Lett. **106**, 082103 (2015).

[18] S. Sengupta, S.Y. Shah, N. Halder, and S. Chakrabarti, Opto-Electronics Rev. **18**, 295 (2010).

[19] M. a. Majid, D.T.D. Childs, K. Kennedy, R. Airey, R. a. Hogg, E. Clarke, P. Spencer, and R. Murray, Appl. Phys. Lett. **99**, 051101 (2011).

[20] M.A. Majid, D.T.D. Childs, H. Shahid, R. Airey, K. Kennedy, R.A. Hogg, E. Clarke, P. Spencer, and R.

Murray, Electron. Lett. **47**, 44 (2011).

[21] M.B. Ward, O.Z. Karimov, D.C. Unitt, Z.L. Yuan, P. See, D.G. Gevaux, A.J. Shields, P. Atkinson, and D.A. Ritchie, Appl. Phys. Lett. **86**, 201111 (2005).

[22] B. Alloing, C. Zinoni, V. Zwiller, L.H. Li, C. Monat, M. Gobet, G. Buchs, A. Fiore, E. Pelucchi, and E. Kapon, Appl. Phys. Lett. **86**, 101908 (2005).

[23] R.P. Mirin, J.P. Ibbetson, J.E. Bowers, and A.C. Gossard, J. Cryst. Growth **175–176**, 696 (1997).

[24] M. V Maximov, A.F. Tsatsul'nikov, B. V Volovik, D.S. Sizov, Y.M. Shernyakov, I.N. Kaiander, A.E. Zhukov, A.R. Kovsh, S.S. Mikhrin, V.M. Ustinov, Z.I. Alferov, R. Heitz, V.A. Shchukin, N.N. Ledentsov, D. Bimberg, Y.G. Musikhin, and W. Neumann, Phys. Rev. B **62**, 16671 (2000).

[25] Y.D. Jang, N.J. Kim, J.S. Yim, D. Lee, S.H. Pyun, W.G. Jeong, and J.W. Jang, Appl. Phys. Lett. **88**, 231907 (2006).

[26] Ł. Dusanowski, M. Syperek, P. Mrowiński, W. Rudno-Rudziński, J. Misiewicz, A. Somers, S. Höfling, M. Kamp, J.P. Reithmaier, and G. Sęk, Appl. Phys. Lett. **105**, 021909 (2014).

[27] M. Benyoucef, M. Yacob, J.P. Reithmaier, J. Kettler, and P. Michler, Appl. Phys. Lett. **103**, 162101 (2013).

[28] K. Takemoto, M. Takatsu, S. Hirose, N. Yokoyama, Y. Sakuma, T. Usuki, T. Miyazawa, and Y. Arakawa, J. Appl. Phys. **101**, 081720 (2007).

[29] M.D. Birowosuto, H. Sumikura, S. Matsuo, H. Taniyama, P.J. van Veldhoven, R. Nötzel, and M. Notomi, Sci. Rep. **2**, 321 (2012).

[30] Ł. Dusanowski, P. Holewa, A. Maryński, A. Musiał, T. Heuser, N. Srocka, D. Quandt, A. Strittmatter, S. Rodt, J. Misiewicz, and S. Reitzenstein, Opt. Express **25**, 340 (2017).

[31] F. Guffarth, R. Heitz, A. Schliwa, O. Stier, N.N. Ledentsov, A.R. Kovsh, V.M. Ustinov, and D. Bimberg, Phys. Rev. B **64**, 85305 (2001).

[32] E. Goldmann, M. Paul, F.F. Krause, K. M??ller, J. Kettler, T. Mehrtens, A. Rosenauer, M. Jetter, P. Michler, and F. Jahnke, Appl. Phys. Lett. **105**, 152102 (2014).

[33] O. Stier, M. Grundmann, and D. Bimberg, Phys. Rev. B **59**, 5688 (1999).

[34] A. Schliwa, G. Hönig, and D. Bimberg, in *Multi-Band Eff. Mass Approx.* (2014), pp. 57–86.

[35] A. Schliwa, M. Winkelnkemper, and D. Bimberg, Phys. Rev. B **79**, 075443 (2009).

[36] F. Olbrich, J. Kettler, M. Bayerbach, M. Paul, J. Höschele, S.L. Portalupi, M. Jetter, and P. Michler, J. Appl. Phys. **121**, 184302 (2017).

[37] M. Paul, J. Kettler, K. Zeuner, C. Clausen, M. Jetter, and P. Michler, Appl. Phys. Lett. **106**, 122105 (2015).

[38] C. Carmesin, F. Olbrich, T. Mehrtens, M. Florian, S. Michael, S. Schreier, C. Nawrath, M. Paul, J. Höschele, B. Gerken, J. Kettler, S.L. Portalupi, M. Jetter, P. Michler, A. Rosenauer, and F. Jahnke, Phys. Rev. B **98**, 125407 (2018).

[39] M. Paul, F. Olbrich, J. Höschele, S. Schreier, J. Kettler, S.L. Portalupi, M. Jetter, and P. Michler, Appl. Phys. Lett. **111**, 033102 (2017).

[40] F. Olbrich, J. Höschele, M. Müller, J. Kettler, S. Luca Portalupi, M. Paul, M. Jetter, and P. Michler, Appl. Phys. Lett. **111**, 133106 (2017).

[41] A. Maryński, P. Mrowiński, K. Ryczko, P. Podemski, K. Gawarecki, A. Musiał, J. Misiewicz, D. Quandt, A. Strittmatter, S. Rodt, S. Reitzenstein, and G. Sęk, Acta Phys. Pol. A **132**, 386 (2017).

[42] P.-I. Schneider, N. Srocka, S. Rodt, L. Zschiedrich, S. Reitzenstein, and S. Burger, Opt. Express **26**, 8479 (2018).

[43] N. Srocka, A. Musiał, P.-I. Schneider, P. Mrowiński, P. Holewa, S. Burger, D. Quandt, A. Strittmatter, S. Rodt, S. Reitzenstein, and G. Sęk, AIP Adv. **8**, 085205 (2018).

[44] M. Sartison, L. Engel, S. Kolatschek, F. Olbrich, C. Nawrath, S. Hepp, M. Jetter, P. Michler, and S.L. Portalupi, Appl. Phys. Lett. **113**, 032103 (2018).

[45] T.F. Kuech, D.J. Wolford, E. Veuhoff, V. Deline, P.M. Mooney, R. Potemski, and J. Bradley, J. Appl. Phys. **62**, 632 (1987).

[46] T.F. Kuech, M.A. Tischler, P. -J. Wang, G. Scilla, R. Potemski, and F. Cardone, Appl. Phys. Lett. **53**, 1317 (1988).

[47] Y. Kodriano, E. Poem, N.H. Lindner, C. Tradonsky, B.D. Gerardot, P.M. Petroff, J.E. Avron, and D. Gershoni, Phys. Rev. B **82**, 155329 (2010).

[48] E.R. Schmidgall, I. Schwartz, L. Gantz, D. Cogan, S. Raindel, and D. Gershoni, Phys. Rev. B **90**, 241411(R) (2014).

[49] M. Wimmer, S. Nair, and J. Shumway, Phys. Rev. B **73**, 165305 (2006).

[50] R.J. Young, R.M. Stevenson, A.J. Shields, P. Atkinson, K. Cooper, D.A. Ritchie, K.M. Groom, a. I. Tartakovskii, and M.S. Skolnick, Phys. Rev. B **72**, 113305 (2005).

[51] M. Bayer, G. Ortner, O. Stern, A. Kuther, A. Gorbunov, A. Forchel, P. Hawrylak, S. Fafard, K. Hinzer, T. Reinecke, S.N. Walck, J.P. Reithmaier, F. Klopf, and F. Schäfer, Phys. Rev. B **65**, 195315 (2002).

[52] S.N. Walck and T. Reinecke, Phys. Rev. B - Condens. Matter Mater. Phys. **57**, 9088 (1998).

[53] J.J. Finley, D. Mowbray, M.S. Skolnick, A. Ashmore, C. Baker, A. Monte, and M. Hopkinson, Phys. Rev. B **66**, 153316 (2002).

[54] A. Musiał, P. Gold, J. Andrzejewski, A. Löffler, J. Misiewicz, S. Höfling, A. Forchel, M. Kamp, G. Sęk, and S. Reitzenstein, Phys. Rev. B **90**, 045430 (2014).

[55] V. Mlinar and A. Zunger, Phys. Rev. B - Condens. Matter Mater. Phys. **79**, 115416 (2009).

[56] R. Seguin, A. Schliwa, S. Rodt, K. Potschke, U.W. Pohl, and D. Bimberg, Phys. Rev. Lett. **95**, 257402 (2005).

[57] P.L. Ardelt, K. Gawarecki, K. Müller, A.M. Waeber, A. Bechtold, K. Oberhofer, J.M. Daniels, F. Klotz, M. Bichler, T. Kuhn, H.J. Krenner, P. Machnikowski, and J.J. Finley, Phys. Rev. Lett. **116**, 077401 (2016).

[58] C.E. Pryor, J. Kim, L.W. Wang, A.J. Williamson, and A. Zunger, J. Appl. Phys. **83**, 2548 (1998).

[59] G. Bester, A. Zunger, X. Wu, and D. Vanderbilt, Phys. Rev. B **74**, 081305 (2006).

[60] T.B. Bahder, Phys. Rev. B **41**, 11992 (1990).

[61] K. Gawarecki, Phys. Rev. B **97**, 235408 (2018).

[62] V. Hernandez, J.E. Roman, and V. Vidal, ACM Trans. Math. Softw. **31**, 351 (2005).

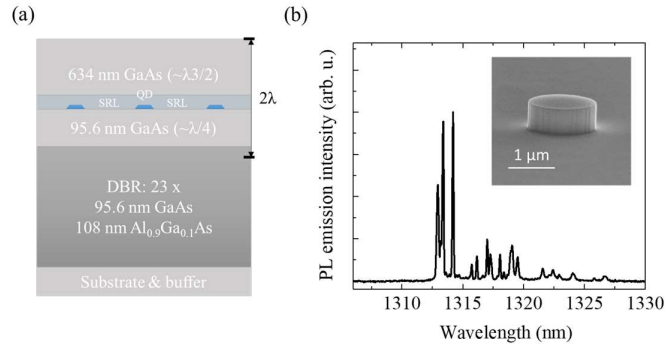

Fig. 1. (a) Sample layout of the investigated QD structure with GaAs substrate and buffer layer, 23 pairs of GaAs/Al$_{0.9}$Ga$_{0.1}$As layers to form the DBR, a single In$_{0.75}$Ga$_{0.25}$As QD layer on wetting layer and embedded in an In$_{0.2}$Ga$_{0.8}$As strain reducing layer (SRL) capped with a thick GaAs layer suitable for the fabrication of nanophotonic mesa structures with a 2-$\lambda$ cavity. (b) High-resolution low-temperature microphotoluminescence spectrum of single QDs located in the 1.5 µm diameter mesa structure obtained under non-resonant excitation. To facilitate single-QD experiments a regular pattern of cylindrical mesas with varying diameters was fabricated using standard electron-beam lithography followed by reactive ion etching (inset: scanning electron microscope image of the investigated mesa structure).

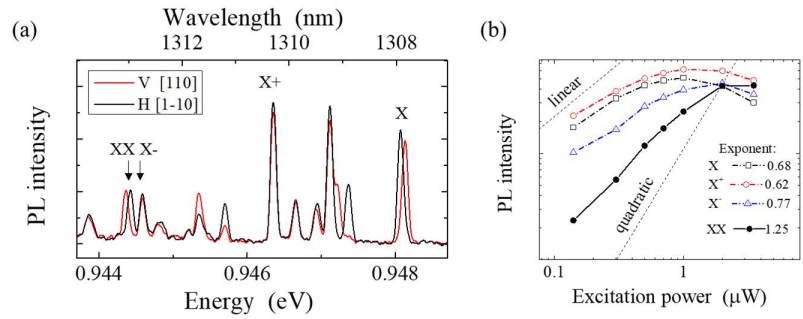

Fig. 2. (a) High-resolution low-temperature microphotoluminescence (µPL) spectra from a single QD for two orthogonal linear polarization directions: along the [1-10] direction (black solid line) and [110] direction (red solid line) with various excitonic complexes marked as X (neutral exciton), XX (biexciton), X$^+$ and X$^-$ (positively and negatively charged excitons). Other spectral features might be related to further excitonic complexes, those including carrier(s) in an excited state. Also, their origin from a second QD located in the same mesa cannot be excluded due to relatively high QD spatial density. (b) Analysis of the µPL emission intensity as a function of excitation power for the emission lines marked in (a).

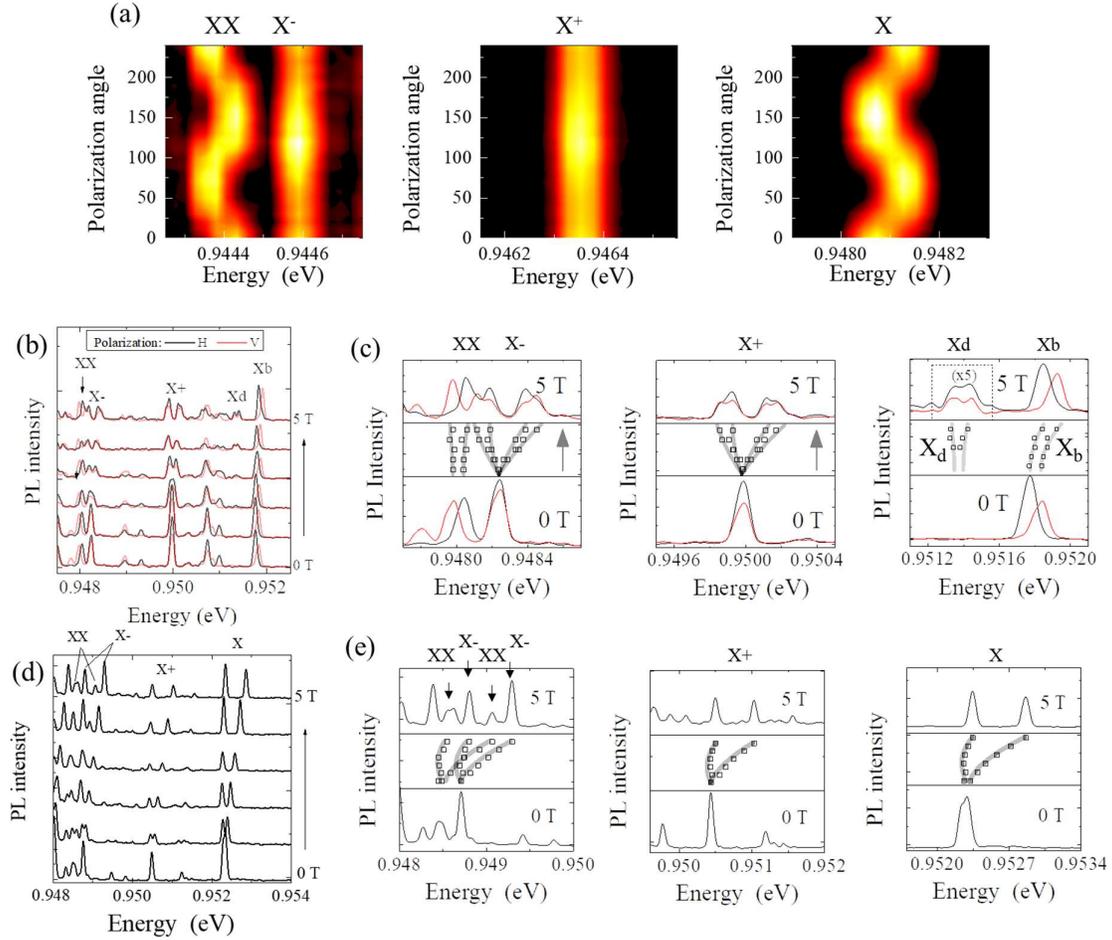

Fig. 3. (a) Polarization-resolved µPL showing the same fine structure splitting (FSS) for X (two bright excitons - $X_b$) and XX with opposite polarizations as well as the lack of the FSS for $X^+$ and $X^-$. (b,c) µPL spectra measured in magnetic field of 0 T and 5 T in Voigt configuration for two orthogonal linear polarizations together with extracted energy dependence of respective emission lines on the strength of the magnetic field showing the same quadruplet splitting of $X^+$ and $X^-$ as well as the dark exciton states $X_d$. The following system parameter values could be determined: $g_{e,\perp} = 0.99$, $g_{h,\perp} = 0.28$ (absolute values), diamagnetic coefficient: $a_{X,\perp} = 2$ µeV/T², exciton fine structure: $\Delta E_{X_b} = 60$ µeV, $\Delta E_{X_d} = 20$ µeV, $\Delta E_{X_b-X_d} = 430$ µeV. (d,e) µPL spectra measured in magnetic field of 0 T and 5 T in Faraday configuration together with extracted energy dependence of the respective emission lines on the strength of the magnetic field, based on which the exciton g-factor (absolute value) equal to 1.75 and diamagnetic coefficient of $a_X = 13$ µeV/T² are obtained.

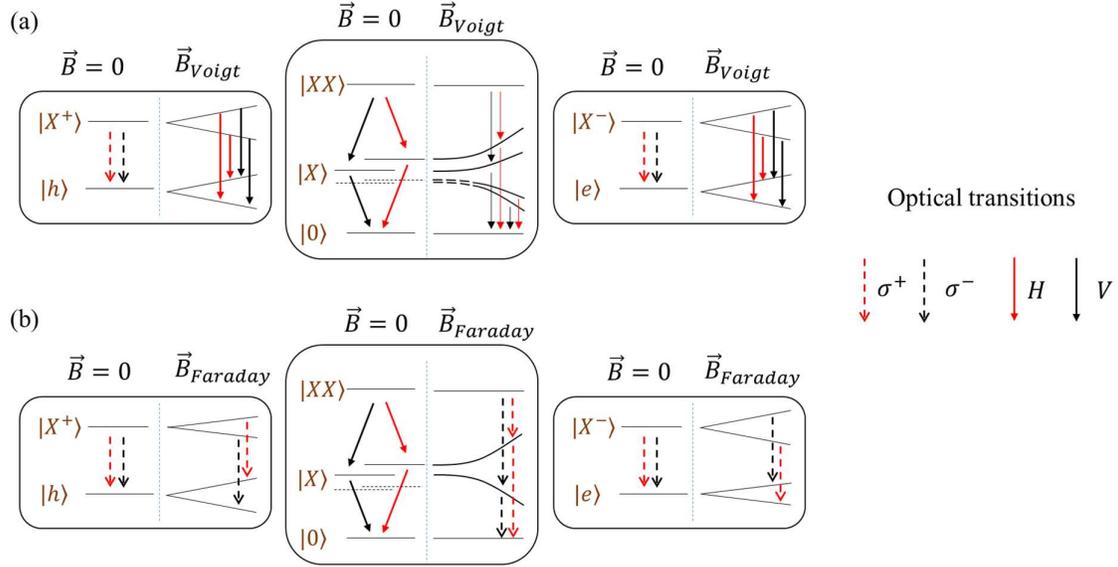

Fig. 4. Energy level scheme for basic excitonic complexes (X – neutral exciton, XX – biexciton, $X^{+/-}$ - positive/negative trion) and its evolution in external magnetic field in Voigt (a) and Faraday (b) configuration due to Zeeman effect (diamagnetic shift neglected for the sake of simplicity). The arrows indicate possible optical transitions and the style of the line corresponds to the polarization of emitted radiation with dotted (solid) lines indicating circular (linear) polarization. H and V corresponds to two orthogonal linear polarizations and $\sigma^{+/-}$ to right and left circular polarization.

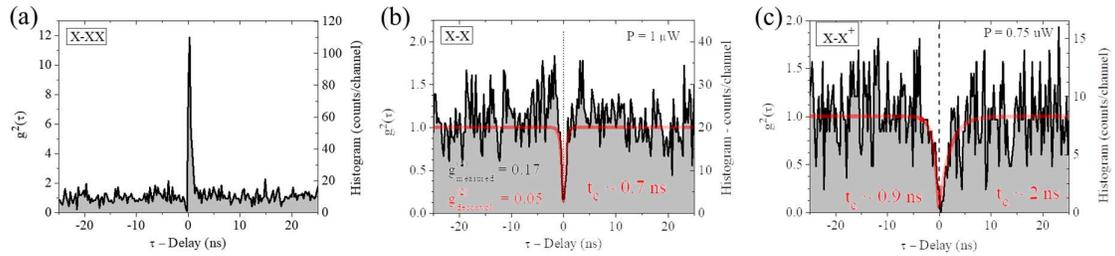

Fig. 5. (a) Cross-correlation of photon emission events for emission lines preliminary identified as XX and X, as well as for (c) X and $X^+$ from a single QD (cf. Fig. 2 and 3); (b) photon auto-correlation for the X emission line measured at 5 K under non-resonant cw excitation: black line – measurement data, red line – (b) symmetric fit and (c) asymmetric fit using $1 - (1 - g^2(0))e^{-\frac{|\tau|}{t_c}}$ function including convolution with the temporal resolution of the experimental setup (80 ps).

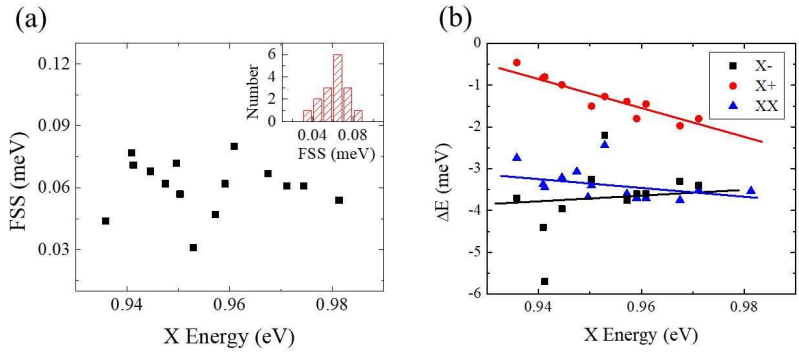

Fig. 6. Summary of the statistical measurements on more than 10 QDs: (a) Fine structure splitting (FSS) of the X (neutral exciton) as a function of X emission energy and FSS histogram (inset). (b) XX (biexciton), $X^+$ (positive trion) and $X^-$ (negative trion) binding energies as a function of X emission energy with solid lines providing guide to the eye.

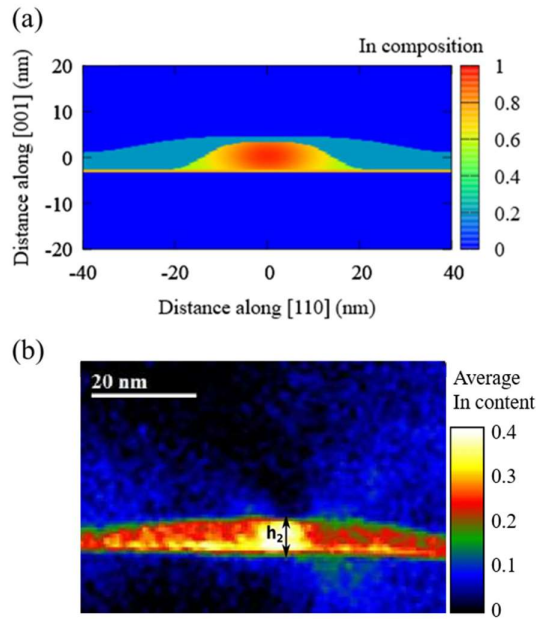

Fig. 7. (a) Applied model structure of a QD. The In composition of $In_xGa_{1-x}As$ is color-coded. (b) Composition Evaluation by Lattice Fringe Analysis (CELFA) image from TEM measurement of the representative InGaAs/GaAs QD taken upon rotation of the sample by 2-4° out of [010] plane with the color scale referring to In content averaged over the sample thickness (see text for interpretation of the signal). Strongly inhomogeneous In distribution within the nanostructure volume is clearly visible.

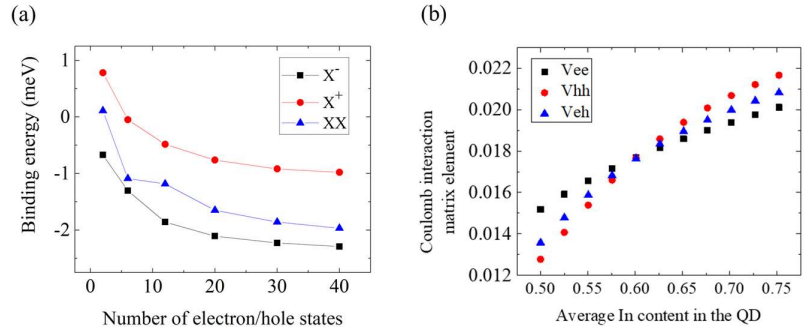

Fig 8. a) Influence of the number of electron/hole basis states included in the configuration interaction calculations on the binding energies of the basic excitonic complexes: positively ($X^+$) and negatively charged ($X^-$) trion and biexciton (XX). b) Coulomb interaction matrix elements for electron-electron $V_{ee}$ (black squares), electron-hole $V_{eh}$ (blue trangles) and hole-hole $V_{hh}$ (red dots) dependence on the average In content in the QD.

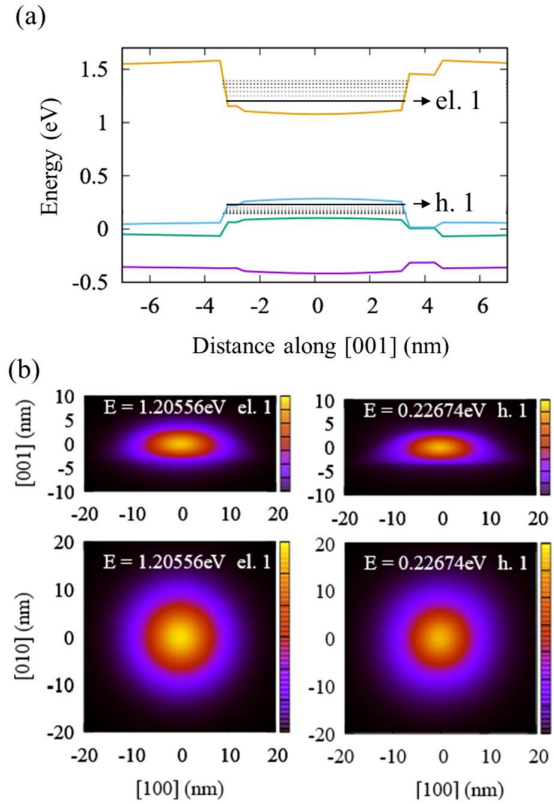

Fig. 9. (a) Single-particle levels (dotted lines) calculated within the 8-band k.p model with the band edge along the growth direction marked for the conduction band (yellow solid line) and for the valence subbands: blue, green and purple for heavy holes, light holes and spin-orbit split-off band, respectively. (b) Probability density of the carrier distribution for the first hole (right column) and electron (left column) levels for cross sections along the growth-direction (upper panel) and in-plane (lower panel) through the center of the QD.

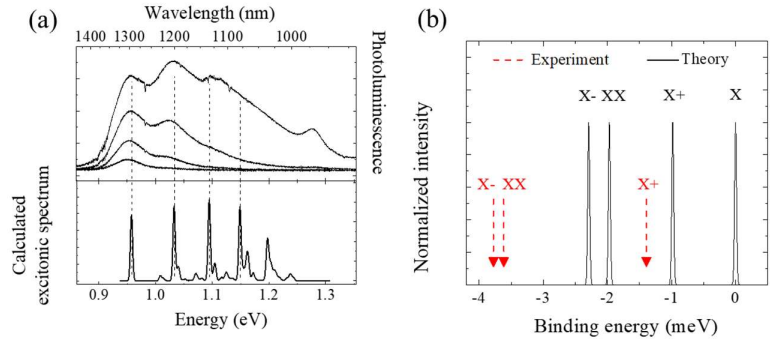

Fig. 10. (a) Photoluminescence spectra of the QD ensemble for increasing excitation power exhibiting the state filling effect with higher energy states clearly resolved (upper panel) compared to the calculated excitonic spectrum (lower panel). (b) Calculated single QD spectrum (black solid line) and experimental values of the binding energies of the excitonic complexes (red arrows). The energy of the neutral exciton from experiment and calculations coincides and it is subtracted from the results.

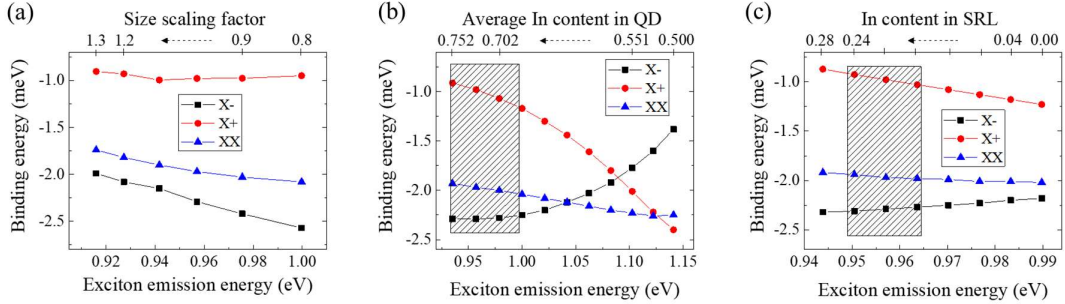

Fig 11. Results from calculations within the 8 band k·p model combined with the configuration interaction method for binding energies of excitonic complexes as a function of: (a) QD size – the height to base size ratio is kept constant so the size scaling factor (SSF) represents the multiplication factor for dimensions in both directions; (b) average In content in the QD and (c) In content in the InGaAs strain reducing layer (SRL). The data are presented in dependence on the exciton emission energy. Nominal fixed parameters: (a) QD $r_b$: 15 nm x SSF, QD h: 6 nm x SSF, average In content in QD: 73 %, In content in SRL: 20 %, (b) QD $r_b$: 15 nm, h: 6 nm, In content in SRL: 20 %, (c) QD $r_b$: 15 nm, h = 6 nm, average In content in QD: 73 %. The marked region in (b) and (c) corresponds to the energy range and SRL composition, respectively of experimental data available in the investigated sample. In the case of (a) the energy ranges are similar for both theoretical calculations and experimental data.